# Gamma Rays From Neutralino Annihilation

Gerard Jungman[a][†] and Marc Kamionkowski[b,c][‡]

[a] *Department of Physics, Syracuse University, Syracuse, NY 13244.*
[b] *School of Natural Sciences, Institute for Advanced Study, Princeton, NJ 08540*
[c] *Department of Physics, Columbia University, New York, NY 10027*

ABSTRACT

We calculate the flux of cosmic gamma rays expected from the annihilation of neutralinos in the Galactic halo. Our calculation of the annihilation cross section to two photons improves the existing calculations by inclusion of exact one-loop diagrams for the amplitudes involving Higgs boson and chargino states as well as those involving fermion and sfermion states. A survey of supersymmetric parameter space shows that numerous models would be observable at the $3\sigma$ level with an air Cerenkov telescope with an exposure of $10^4$ m$^2$ yr.

25 October 1994

[†] jungman@npac.syr.edu
[‡] kamion@phys.columbia.edu

# 1. INTRODUCTION

For quite some time, it has been known that only about a tenth of the mass of most galaxies, including our own, is luminous, and that the rest is composed of some sort of dark matter [1]. The nature of this nonluminous material is unknown, although there are quite convincing arguments that it must be non-baryonic. One of the most promising of many candidates for the dark matter is the neutralino [2][3], a linear combination of the supersymmetric partners of the photon, $Z^0$, and Higgs bosons. It has been suggested by numerous authors that if neutralinos populate the Galactic halo, then monoenergetic gamma rays produced by neutralino annihilation in the halo could provide a plausible avenue toward discovery of such dark-matter particles [4]. In this paper we re-examine this proposal.

The goal of this work is to provide results for the cross section for annihilation of neutralinos to two photons which include all of the contributions at one loop to the amplitude for a neutralino in any given minimal supersymmetric extension of the standard model. We generalize previous calculations of the amplitude for annihilation through quark-squark loops [5][6][7] and Higgs-chargino loops [5] to arbitrary neutralino and squark masses and compositions. We also include the recent approximate calculation of Bergstrom and Kaplan [8] of annihilation through $W^\pm$-boson–chargino loops and improve it by including subleading logarithmic terms.

We then use these expressions in a survey of supersymmetric parameter space to assess the possibility of discovering dark-matter neutralinos in the Galaxy via observation of cosmic gamma rays produced by neutralino annihilation.

In the following, we will present our result for the cross section, discussing the importance of the various contributions. Then we will give an estimate of the signal rates which are implied.

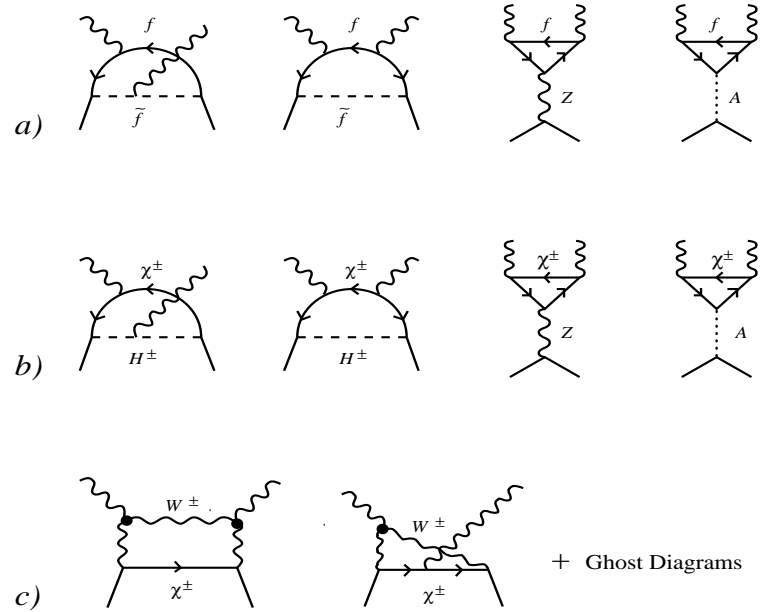

FIG. 1. Feynman diagrams for neutralino annihilation to photons.

# 2. CROSS SECTION

The Feynman diagrams for annihilation to two photons are shown in Fig. 1. The diagrams fall into three categories. The similarity between the diagrams of classes a) and b) indicates that the corresponding amplitudes can be written in terms of the same basic functions arising from the loop integrations. Furthermore, these functions are precisely those which appear in the calculation of the cross section for neutralino annihilation to two gluons presented in Ref. [9]. The third class of diagrams are those with $W$ bosons in the loop, and those ghost diagrams which are related to them; this gauge invariant set of diagrams has been discussed in Ref. [8]. By a choice of non-linear gauge, the calculation was reduced significantly. In the limit $m_{\chi^\pm} \gtrsim m_{\chi^0}$ (where $m_{\chi^\pm}$ and $m_{\chi^0}$ are masses of the chargino $\chi^\pm$ and neutralino $\chi^0$, respectively), which is always appropriate when considering the neutralino as the lightest supersymmetric particle, the amplitude reduces to a single three-point integral, as given in Ref. [8].

Neutralinos in the halo move with velocities negligible compared with the speed of light, so annihilation occurs in the $s$ wave only. Therefore, the amplitudes depend only on the outgoing photon momenta and polarizations, and the amplitudes can be written in the form

$$\mathcal{A} = \frac{e^2}{4\pi^2} \epsilon(k_1, k_2, \epsilon_1, \epsilon_2) \tilde{\mathcal{A}}, \tag{2.1}$$

where $k_i$ and $\epsilon_i$ are the momenta and polarizations of the outgoing photon pair. The total amplitude will be a sum of three parts to be discussed in turn below, $\mathcal{A} = \mathcal{A}_{f\bar{f}} + \mathcal{A}_{HC} + \mathcal{A}_W$. Given this amplitude, the cross section is

$$\sigma_{\gamma\gamma} v = \frac{\alpha^2 m_\chi^2}{16\pi^3} |\tilde{\mathcal{A}}|^2. \tag{2.2}$$

First, consider the amplitudes related to the two gluon amplitudes, i.e., those coming from fermion-sfermion loop diagrams shown in class a) in Fig. 1. Define the following functions, arising from the loop integrals:

$$F_I(a, b, S, D) = \frac{1}{1+a-b} \left(\frac{1}{2}\right) \left[Sb + D\sqrt{ab}\right], \tag{2.3}$$

$$T(c, \kappa_A, \kappa_Z) = 2\sqrt{c} \frac{\kappa_A \, g_{A\chi\chi}}{4 - m_\chi^2/m_A^2} + c \frac{m_\chi^2}{m_Z^2} \kappa_Z \, g_{Z\chi\chi}, \tag{2.4}$$

where $m_A$ and $m_Z$ are the masses of the pseudoscalar-Higgs and $Z$ bosons, $g_{A\chi\chi}$ and $g_{Z\chi\chi}$ are the couplings of neutralinos to the pseudoscalar-Higgs and $Z$ bosons [9], and

$$F(a, b, S, D) = -\frac{1}{2} \int_0^1 dx \left\{ \frac{S}{x} \ln \left| \frac{x^2 a + x(b-1-a) + 1}{-x^2 a + x(b-1+a) + 1} \right| \right.$$
$$+ \frac{Sb + D\sqrt{ab}}{1+a-b} \left(\frac{1}{1-x} + \frac{1}{1+x}\right) \ln \left| \frac{x^2 a + x(b-a-1) + 1}{b + a(1-x^2)} \right|$$
$$+ \frac{1}{1-b+xa} \left[Sb \left(\frac{1}{x} + \frac{1}{1-x}\right) + D \frac{\sqrt{ab}}{1-x}\right] \ln \left| \frac{b}{x^2 a - x(a+b-1) - 1} \right|$$
$$\left. + \frac{1}{b-1+ax} \left[Sb \left(\frac{1}{x} - \frac{1}{1+x}\right) - D \frac{\sqrt{ab}}{x+1}\right] \ln \left| \frac{b}{x^2 a + x(b-1-a) + 1} \right| \right\}. \tag{2.5}$$

Note that this expression for $F(a, b, S, D)$ corrects a typographical error in Ref. [9].

The amplitude in terms of these functions is

$$\mathrm{Re}\tilde{\mathcal{A}}_{f\bar{f}} = \sum_f c_f Q_f^2 \left\{ 2I\left(\frac{m_f^2}{m_\chi^2}\right) \frac{1}{m_\chi^2} T\left(\frac{m_f^2}{m_\chi^2}, g_{Aff}, \frac{g}{\cos\theta_W}\right) \right.$$
$$\left. + \sum_{\tilde{f}} \frac{1}{m_\chi^2} F\left(\frac{m_\chi^2}{m_{\tilde{f}}^2}, \frac{m_f^2}{m_{\tilde{f}}^2}, S_{f\bar{f}}, D_{f\bar{f}}\right) \right\},$$
$$\mathrm{Im}\tilde{\mathcal{A}}_{f\bar{f}} = -\pi \sum_f c_f Q_f^2 \theta(m_\chi^2 - m_f^2) \ln \frac{1+\beta_f}{1-\beta_f} \left\{ -T\left(\frac{m_f^2}{m_\chi^2}, g_{Aff}, \frac{g}{\cos\theta_W}\right) \right.$$
$$\left. + \frac{1}{m_\chi^2} \sum_{\tilde{f}} F_I\left(\frac{m_\chi^2}{m_{\tilde{f}}^2}, \frac{m_f^2}{m_{\tilde{f}}^2}, S_{f\bar{f}}, D_{f\bar{f}}\right) \right\}, \tag{2.6}$$

where the sum on $f$ is over quarks and leptons, and the sum on $\tilde{f}$ is over the squarks and sleptons. Here, $g_{Aff}$ are the pseudoscalar-Higgs–fermion couplings [9], $\beta_f = (1 - m_f^2/m_\chi^2)^{1/2}$, $Q_f$ is the electric charge of $f$, $c_f$ is a color factor which equals 3 for quarks and 1 for leptons, and $\theta(x)$ is the Heaviside step function. The function $I(x)$ is given in Eq. (2.14) of Ref. [9]. The couplings of the fermions and sfermions are defined by $S_{f\bar{f}} = A_{f\bar{f}}^2 + B_{f\bar{f}}^2$ and $D_{f\bar{f}} = A_{f\bar{f}}^2 - B_{f\bar{f}}^2$, with

$$A_{f\bar{f}} = \frac{1}{2} \left(X'_{f\bar{f}0} + W'_{f\bar{f}0}\right),$$
$$B_{f\bar{f}} = \frac{1}{2} \left(X'_{f\bar{f}0} - W'_{f\bar{f}0}\right). \tag{2.7}$$

The fundamental couplings $X'$ and $W'$ are the couplings of left-handed and right-handed fermions, respectively, to sfermions and neutralinos, as defined in Refs. [9] and [10].

Next consider the amplitude from diagrams involving intermediate charged-Higgs bosons $H^\pm$ and charginos, i.e., those in class b). We find

$$\mathrm{Re}\tilde{\mathcal{A}}_{HC} = \sum_{\chi_m^\pm} \left\{ 2I\left(\frac{m_{\chi_m^\pm}^2}{m_\chi^2}\right) T\left(\frac{m_{\chi_m^\pm}^2}{m_\chi^2}, h_{A\chi_m^\pm\chi_m^\pm}, \delta_m\right) \right.$$
$$\left. + F\left(\frac{m_\chi^2}{m_{H^\pm}^2}, \frac{m_{\chi_m^\pm}^2}{m_{H^\pm}^2}, S_m, D_m\right) \right\}, \tag{2.8}$$

where the sum is over the two charginos. Here $\delta_m = (g/\cos\theta_W)(O'^L_{mm} - O'^R_{mm})$ is the coupling of the $Z$ to a chargino pair, and $h_{A\chi^\pm_m\chi^\pm_m} = g(\sin\beta Q_{mm} + \cos\beta S_{mm})$ is the coupling of the $A$ boson to a chargino pair. The couplings of the charginos to neutralinos are defined by

$$S_m = \frac{1}{2}\left[\left(Q'^L_{0m}\right)^2 + \left(Q'^R_{0m}\right)^2\right],$$
$$D_m = Q'^L_{0m} Q'^R_{0m}. \qquad (2.9)$$

Definitions and more details of the couplings $O'$ and $Q'$ can be found in Refs. [2] and [10].

Next consider the amplitude from diagrams with intermediate $W$ bosons, i.e., those in class c). In Ref. [8], the imaginary part of this amplitude was calculated exactly,

$$\text{Im}\tilde{\mathcal{A}}_W = \pi \sum_{\chi^\pm_m} \theta(m^2_\chi - m^2_W) C_m \beta^2_W \ln\frac{1+\beta_W}{1-\beta_W}, \qquad (2.10)$$

where $\beta_W = (1 - m^2_W/m^2_\chi)^{1/2}$. The coupling is given by

$$C_m = \frac{g^2}{2\sqrt{2}} \frac{4}{m^2_{\chi^\pm_m}} \left[\left(O^L_{0m}\right)^2 + \left(O^R_{0m}\right)^2\right], \qquad (2.11)$$

where $O^L_{0m}$ and $O^R_{0m}$ are given in Refs. [2] and [10]. The real part of the amplitude was evaluated in a leading-logarithm limit in Ref. [8], using dispersion relations, giving a result proportional to $\ln^2(a)$. As pointed out there, it is an excellent approximation to contract the chargino propagator to a point, corresponding to the limit $m_{\chi^\pm} \gtrsim m_\chi$. Evaluating the real part of the subsequent three-point amplitude, we find

$$\text{Re}\tilde{\mathcal{A}}_W = -2\sum_{\chi^+_m} C_m \left\{-\frac{3}{4} - \frac{1}{2}\ln\frac{m^2_W}{4m^2_\chi} + \frac{1}{2}\ln\frac{4m^2_\chi}{\mu^2} + \frac{1}{4}B_1\left(\frac{m^2_W}{4m^2_\chi}\right) - B_2\left(\frac{m^2_W}{4m^2_\chi}\right)\right\}, \qquad (2.12)$$

where

$$B_1(a) = \int_0^1 \frac{dx}{x}\left[1 + \frac{a-x(1-x)}{x}\ln\left|\frac{a-x(1-x)}{a}\right|\right],$$
$$B_2(a) = \int_0^1 \frac{dx}{x}[a + x(1-x)]\ln\left|\frac{a-x(1-x)}{a}\right|, \qquad (2.13)$$

and $\mu$ is the renormalization point; note that $\mu$ appears because of the contraction of the chargino propagator. We choose $\mu = m_W$, specifying the running coupling at the electroweak scale; this gives a subleading logarithmic contribution to the amplitude. It is very useful to have an asymptotic expansion for this result, extending the leading-logarithm expansion of Ref. [8] into the subleading terms. We find

$$B_1(a) \sim \begin{cases} -\frac{1}{2}\ln^2 a + 1 + \frac{\pi^2}{3} & a \to 0, \\ 1 + \frac{1}{6a} & a \to \infty, \end{cases}$$
$$B_2(a) \sim \begin{cases} -\frac{1}{2}\ln a - 1 & a \to 0, \\ -\frac{1}{2} - \frac{1}{8a} & a \to \infty. \end{cases} \qquad (2.14)$$

The leading logarithmic behaviour of this amplitude agrees with that obtained in Ref. [8]. It should be pointed out that these expressions are not strictly valid for $m_\chi < m_W$. However, since we are interested in high energy gamma rays, this case is not of interest to us; we henceforth assume that $m_\chi > m_W$.

Now we can consider the behaviours of each of these terms. The size of the fermion/sfermion loop contribution is sensitive to the gaugino fraction of the neutralino and to the masses of the sfermions; a sufficiently Higgsino-like composition of the neutralino will suppress these contributions. As pointed out by Bergstrom and Kaplan [8], annihilation through loops containing $W$ bosons [class c)] may be significant.

The Higgs-chargino loop contribution can be important for several reasons. First note that the couplings which appear in these diagrams are all naturally of the order of the gauge couplings, so that there is no special suppression, as can sometimes occur in couplings of neutralinos to fermions and sfermions. Furthermore, the masses of the intermediate particles are comparable, and this matching of scales provides some enhancement of the loop integral. In fact, it is important to note that the amplitude as written contains a pole at the point $b = 1 + a$, or $m^2_\chi + m^2_{H^\pm} = m^2_{\chi^\pm}$. This divergence occurs because we have ignored the widths of the intermediate particles. In actuality these widths are large; for chargino masses greater than 100 GeV the width of the chargino is approximately $\Gamma_{\chi^\pm} \simeq 0.1 m_{\chi^\pm}$. So the pole is spurious, and the amplitude must be mollified in the region $b \simeq 1 + a$. Note that previous treatments of this contribution [5] will not be reliable in the common case that the masses of the particles are comparable.

The $W$-loop contribution is almost always important in the case that the neutralino is heavier than the $W$ and primarily Higgsino [8]. This is because this contribution depends most strongly on the lightest-chargino mass and this is usually not too much larger than the neutralino mass.

## 3. SIGNAL RATES

We follow Urban et al. [11] and consider the signal from pointed observation of the Galactic center with an atmospheric Cerenkov telescope (ACT). This is potentially a promising method for observation of high-energy gamma rays from neutralino annihilation.

The flux of gamma rays from neutralino annihilation, from a window of solid angle $\Delta\Omega$ aimed at the Galactic center, may be written [4],

$$\phi_\gamma \simeq 2 \times 10^{-11} \text{ cm}^{-2} \text{ sec}^{-1} (\rho_\chi^{0.4})^2 f_{\text{halo}} \frac{[(\sigma_{\gamma\gamma} v)/10^{-30} \text{ cm}^3 \text{ sec}^{-1}]}{(m_\chi/10 \text{ GeV})^2} (\Delta\Omega/\text{sr}), \quad (3.1)$$

where $\rho_\chi^{0.4}$ is the local halo density in units of 0.4 GeV cm$^{-3}$, and $0.3 \lesssim f_{\text{halo}} \lesssim 2$ is a parameter which reflects uncertainty in modeling the galactic halo. Eq. (3.1) is obtained assuming an isothermal halo with a density profile suitable for accounting for the observed rotation curve. If the cosmological neutralino density is too small to account for the halo dark matter, neutralinos should still gather in galactic halos, and the annihilation cross section in this case should generally be larger than that in the case where neutralinos are the dark matter. Therefore, there may be an observable gamma-ray signature of neutralinos even if they exist but are too few to account for all the halo dark matter. To account for this, we take the halo density (for $\Omega_\chi h^2 \lesssim 1$ as required by the age-of-the-Universe constraint) to be $\Omega_\chi h^2/0.25$, where $\Omega_\chi$ is the cosmological neutralino abundance, and $h$ is the Hubble parameter in units of 100 km sec$^{-1}$ Mpc$^{-1}$. We note that our results will not depend sensitively on this prescription.

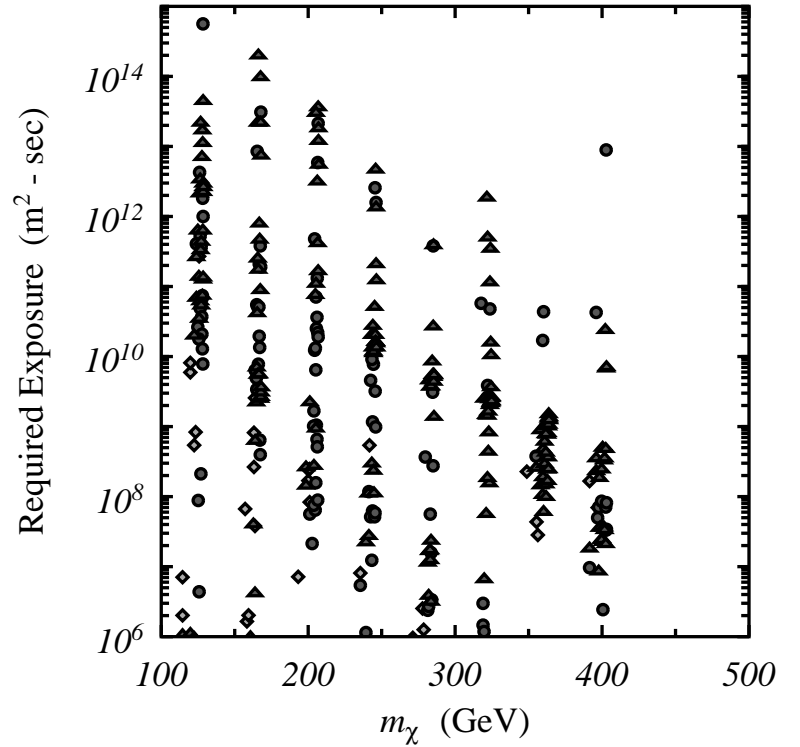

FIG. 2. Minimum exposure required for a $3\sigma$ detection of gamma rays from neutralino annihilation in the Galactic center, versus mass of the neutralino, for the survey of supersymmetric parameter space discussed in the text. The symbols indicate which of the three amplitude contributions dominates the cross section; triangles indicate that the fermion-sfermion diagrams dominate, diamonds indicate that the $W$ diagrams dominate, and circles indicate that the Higgs-chargino diagrams dominate.

The natural width of the two-photon peak is small, and the background below the peak is controlled by the instrumental resolution. The most important background for energies below 1 TeV comes from misidentified charged particles [11]; the background from diffuse cosmic gamma rays [6] contributes only slightly to the total background in this regime. Following Ref. [11], we

7  8

find a background flux of

$$\phi_b = (1.2 \times 10^{-5} \, \text{sec}^{-1} \text{m}^{-2}) \frac{\Delta E}{100 \, \text{GeV}} \left[ \frac{\Delta \Omega}{10^{-3} \text{sr}} \right] \left[ \frac{m_\chi}{100 \, \text{GeV}} \right]^{-p}, \qquad (3.2)$$

where $\Delta E$ is the energy resolution on the peak, and $p \simeq 3.3$ is the spectral index for the background, in this case dominated by misidentified electrons [11].

In Fig. 2 we plot the exposure required for a $3\sigma$ detection of gamma rays from neutralino annihilation in the Galactic center for a variety of models. The parameter ranges which generated these models were taken to be 100 GeV $< M_2 <$ 800 GeV, 200 GeV $< \mu <$ 800 GeV, $2 < \tan\beta < 20$, 300 GeV $< m_A <$ 600 GeV, and 200 GeV $< m_{\tilde{q}} <$ 800 GeV. Here $M_2$ and $\mu$ are the gaugino mass parameters, $m_A$ is the mass of the pseudoscalar Higgs particle, and $m_{\tilde{q}}$ is a common squark mass parameter (which differs from the actual squark masses due to mixing terms, which were included). The grand unification condition was assumed. Models were cut from the plot if they violated known bounds from $e^+e^-$ physics, if they gave Higgs masses in volation of current limits, or if they were inconsistent as models for neutralino dark matter (for example, we obviously require that the lightest supersymmetric particle is a neutralino).

Triangles, circles, and diamonds indicate models where the fermion-sfermion, Higgs-chargino, and $W$-boson diagrams, respectively, dominate the amplitude. According to Fig. 2, the fermion-sfermion diagrams are most often important, but there are indeed regions of parameter space where the Higgs-chargino and $W$ loops are dominant. These results indicate that numerous supersymmetric models could be probed by an atmospheric Cerenkov detector with an area $\mathcal{O}(10^4 \, \text{m}^2)$.

## 4. ACKNOWLEDGMENTS

G.J. thanks J. Buckley for conversations about high-energy gamma rays. M.K. acknowledges the hospitality of the Theory Group at CERN where part of this work was completed. M.K. was supported at the I.A.S. by the W. M. Keck Foundation and at Columbia University by the U.S. Department of Energy under contract DE-FG02-92ER40699. G.J. was supported by the U.S. D.O.E. under contract DE-FG02-85ER40231.

## References

[1] For recent reviews of dark matter and its detection, see V. Trimble, *Ann. Rev. Astron. Astrophys.* **25**, 425 (1989); J. R. Primack, B. Sadoulet, and D. Seckel, *Ann. Rev. Nucl. Part. Sci.* **B38**, 751 (1988); *Dark Matter in the Universe*, eds. J. Kormendy and G. Knapp (Reidel, Dordrecht, 1989).

[2] H. E. Haber and G. L. Kane, *Phys. Rep.* **117**, 75 (1985).

[3] J. Ellis, J. S. Hagelin, D. V. Nanopoulos, K. A. Olive, and M. Srednicki, *Nucl. Phys.* **B238**, 453 (1984); K. Griest, M. Kamionkowski, and M. S. Turner, *Phys. Rev. D* **41**, 3565 (1990); M. Drees and M. M. Nojiri, *Phys. Rev. D* **47**, 376 (1993); K. A. Olive and M. Srednicki, *Phys. Lett.* **B230**, 78 (1989); K. A. Olive and M. Srednicki, *Nucl. Phys.* **B355**, 208 (1991); J. McDonald, K. A. Olive, and M. Srednicki, *Phys. Lett.* **B283**, 80 (1992); K. Griest, *Phys. Rev. D* **38**, 2357 (1988); FERMILAB-Pub-89/139-A (E); *Phys. Rev. Lett.* **61**, 666 (1988).

[4] M. Kamionkowski, to appear in *The Gamma-Ray Sky with Compton GRO and SIGMA*, proceedings of the Winter School, Les Houches, France, Jan 5-Feb 4, 1994.

[5] G. F. Giudice and K. Griest, *Phys. Rev. D* **40**, 2549 (1989).

[6] S. Rudaz, *Phys. Rev. D* **39**, 3549 (1989).

[7] L. Bergstrom, *Phys. Lett.* **B225**, 372 (1989).

[8] L. Bergstrom and J. Kaplan, preprint USITP-94-03, hep-ph/9403239 (1994).

[9] M. Drees, G. Jungman, M. Kamionkowski, and M. Nojiri, *Phys. Rev. D* **49**, 636 (1994).

[10] K. Griest, G. Jungman, and M. Kamionkowski, to be submitted to *Physics Reports*.

[11] M. Urban et al., *Phys. Lett.* **B293**, 149 (1992).